\RequirePackage{fix-cm}
\RequirePackage{graphicx}
\documentclass[twocolumn,draft,epjc3]{svjour3} 

\journalname{Eur. Phys. J. C}

\usepackage[utf8]{inputenc}

\usepackage{xcolor}
\usepackage{amsmath,amssymb,amsfonts,epsfig}
\usepackage{comment}

\usepackage[normalem]{ulem} 

\def \trace{\text{Tr}}

\allowdisplaybreaks[1]
\begin{document}

\setlength{\parindent}{0pt}


\title{Lepton universality and lepton flavor conservation tests  with dineutrino modes}

\author{Rigo Bause\thanksref{e1} \and Hector Gisbert\thanksref{e2}  \and Marcel Golz\thanksref{e3}  \and Gudrun Hiller\thanksref{e4} }

\institute{Fakult\"at f\"ur Physik, TU Dortmund, Otto-Hahn-Str.\,4, D-44221 Dortmund, Germany}

\date{Received: date / Accepted: date}

\thankstext{e1}{rigo.bause@tu-dortmund.de}
\thankstext{e2}{hector.gisbert@tu-dortmund.de}
\thankstext{e3}{marcel.golz@tu-dortmund.de}
\thankstext{e4}{ghiller@physik.uni-dortmund.de}

\maketitle

\begin{abstract} 
$SU(2)_L$-invariance links charged dilepton \linebreak and dineutrino couplings. 
Phenomenological implications 
are worked out  for  flavor changing neutral currents
involving strange, charm, beauty and top
quark transitions in a model-independent way.
We put forward novel tests of lepton universality  and  lepton flavor conservation in  $|\Delta c|= |\Delta u|=1$  and
 $|\Delta b|= |\Delta q|=1$, $q=d,s$, transitions suitable for  the experiments  Belle II and BES III.
Single top production plus dileptons uniquely probes
semileptonic four-fermion  $|\Delta t|= |\Delta q'|=1$, $q'=u,c$, couplings and further study at the LHC is encouraged.
Tests with  single top production associated with dileptons and missing energy are suitable for study at future $e^+ e^-$ or muon colliders.

\end{abstract}

\section{Introduction}

Gauge and approximate flavor symmetries of the standard model (SM) provide powerful means to probe new physics (NP). In this work we exploit the $SU(2)_L$-link between left-handed charged lepton and neutrino couplings, 
to put quantitatively lepton universality (LU) and charged lepton flavor conservation (cLFC) to the test with dineutrino observables.
Schematically,
\begin{align}  \nonumber
&O (\nu \bar \nu)=\sum_{i,j} O (\nu_i  \bar \nu_j) =\sum_{i,j} O  (\ell^-_i  \ell^+_j)  \\
&= O  (e^- e^+)  + O  (\mu^- \mu^+) +O  (\tau^- \tau^+) +O  (e^- \tau^+)  + \ldots 
\nonumber
\end{align}
where we used that the neutrino flavors $i,j$ are not reconstructed and the ellipses stand for the remaining charged lepton flavor violating (cLFV) terms.
Apparently observables $O (\nu \bar \nu)$ can be bounded by a sum of lepton-flavor specific data.
Assuming cLFC, only same-flavor charged dilepton observables contribute.
Assuming in addition universality, the strongest limit among the dielectron, dimuon and ditau
observables dictates the bound.
Confronting data on dineutrino observables  to  charged dilepton ones hence probes lepton flavor.

Somewhat paradoxically,  the method works despite the fact that in dineutrino observables lepton flavor is not determined.
The method is independent of the neutrino mixing matrix and holds for NP contributions from above the weak scale.
 The relation works of course also the other way around: dineutrino limits bounding charged dilepton couplings \cite{Grossman:1995gt}.
 In the end it is a matter of available data which direction is informative on BSM physics.
 Given the huge  improvements over the past years  this analysis is timely.
 With the flavor anomalies challenging LU~\cite{Hiller:2014qzg,Ciezarek:2017yzh}
such quantitative data-driven tests of lepton flavor are  important and of recent interest.

We consider observables involving quark flavor changing neutral currents (FCNCs) and dineutrinos, but note that the method works analogously for flavor  conserving quark transitions.
We furthermore employ the standard model effective theory (SMEFT), which
provides a model-independent parametrization of NP in  terms of higher dimensional operators composed out of SM degrees of freedom and which respect SM gauge and Lorentz invariance. 
SM gauge symmetry also dictates a relation between the weak isospin partners of the left-handed $SU(2)_L$ doublet quarks.
The situation is very different for FCNC processes involving up-type quarks and down-type quarks due to a
 strong Glashow-Iliopoulos-Maiani (GIM)-suppression of SM amplitudes in the former \cite{Eilam:1990zc,Burdman:2001tf}, together with weaker experimental constraints, compared to the latter.
 As such, rare kaon and $b$-decay data constrain charm and top FCNCs, respectively, and not the other way around.
The other important consequence of the GIM-mechanism is that signals  of up-type dineutrino FCNCs  are automatically signals of NP. It is a central point of this work  that, once observed, 
the rate at which these missing energy processes occur is  informative on the violation of LU and cLFC, complementing tests with lepton-specific ratio-
 ${\cal{O}}(\mu^+ \mu^-)/{\cal{O}}(e^+  e^-)$  a la  $R_K$~\cite{Hiller:2014qzg}.

The aim of this work is twofold: Firstly, to present a global perspective of LU and cLFC tests using dineutrino couplings in quark FCNCs, 
 by discussing jointly different flavor sectors,
 strange, charm, beauty and top. Secondly,  we work out implications from rare kaon decays and investigate opportunities with  top FCNCs to
 probe for universality violation and beyond.
  While the connection between lepton flavor violation and  dineutrino branching ratios has been discussed for  kaons \cite{Grossman:2003rw} and $B$-decays, {\it e.g.}~\cite{Buras:2014fpa} and references therein, the  actual relation we obtain, \eqref{eq:super}, and its use in concrete tests of lepton flavor universality is new.

The paper is organized as follows: In Sec.~\ref{sec:SU2} we give the effective theory framework and general relations between neutrino and charged lepton couplings.
We analyze NP implications for first and second generation quarks, strangeness and charm, in Sec.~\ref{sec:12}, and for
processes involving third generation quarks in  Sec.~\ref{sec:3}.
In Sec.~\ref{sec:con} we summarize.

\section{Model-independent relation and lepton flavor \label{sec:SU2}}

Consider the $SU(3)_C \times SU(2)_L \times U(1)_Y$-invariant effective theory with semileptonic (axial-) vector four-fermion operators induced by NP at a scale sufficiently separated from the weak scale $v=(\sqrt{2}\,G_F)^{-1/2} \simeq 246\,$GeV at lowest order~\cite{Grzadkowski:2010es}, with Wilson coefficients $C^{(1)}_{\ell q}$,\, $C^{(3)}_{\ell q}$,\, $C_{\ell u}$,\, $C_{\ell d}$,

\begin{align}\label{eq:ops} 
\begin{split}
{\mathcal{L}}_{\text{eff}} & \supset \frac{C^{(1)}_{\ell q}}{v^2} \bar Q \gamma_\mu Q \,\bar L \gamma^\mu L +\frac{C^{(3)}_{\ell q}}{v^2} 
\bar Q \gamma_\mu  \tau^a Q \,\bar L \gamma^\mu \tau^a L   \\
&+\frac{C_{\ell u}}{v^2}  \bar U \gamma_\mu U \,\bar L \gamma^\mu L +
\frac{C_{\ell d }}{v^2} \bar D \gamma_\mu D \,\bar L \gamma^\mu L \,. 
\end{split}
\end{align}

\begin{figure}
    \centering
    \includegraphics[width=0.2\textwidth]{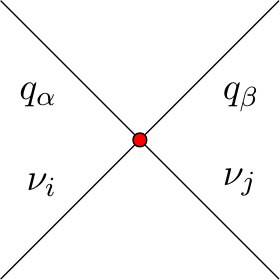}
    \includegraphics[width=0.13\textwidth]{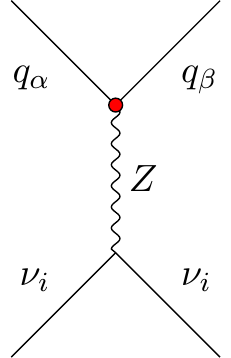}
    \includegraphics[width=0.13\textwidth]{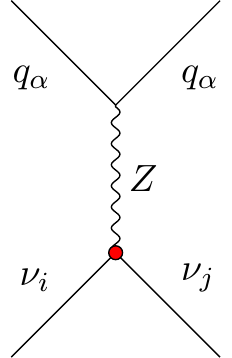}
    \caption{Contributions of four-fermion operators \eqref{eq:ops}, left-most diagram, and subleading ones involving $Z$-exchange,  see text, to quark $q$ processes into neutrinos, contained in the left-handed lepton doublet,
    with flavor indices $\alpha, \beta,i,j$.
    Operators are denoted by a blob.}
    \label{fig:ops}
\end{figure}

Here, $\tau^a$ are Pauli-matrices, $Q$ and $L$ denote 
left-handed quark and lepton $SU(2)_L$-doublets 
and $U,D$ stand for right-handed
up-singlet, down-singlet quarks,
respectively, with flavor indices suppressed for brevity. 
No further dimension six four-fermion operators exist that contribute 
at lowest order to dineutrino modes assuming only SM-like light neutrinos. 
Operators with dimension larger than six contribute at least with relative suppression by two orders of the ratio of weak to NP scales. 
Operators with quarks or leptons with two Higgs fields $\Phi$ and a covariant de\-ri\-vative $D^\mu$, 
$\bar Q \gamma_\mu Q \,\Phi^\dagger D^\mu \Phi$,\, $\bar Q \gamma_\mu \tau^a Q \,\Phi^\dagger D^\mu \tau^a \Phi$,\,
$\bar U \gamma_\mu U\, \Phi^\dagger D^\mu \Phi$,\,
$\bar D \gamma_\mu D\, \Phi^\dagger D^\mu \Phi$
and $\bar L \gamma_\mu L \,\Phi^\dagger D^\mu \Phi$, 
$\bar L \gamma_\mu \tau^a L \,\Phi^\dagger D^\mu \tau^a \Phi$
induce $Z$-penguins at tree level, see Fig.~\ref{fig:ops},
-- the lepton ones conserve quark flavor, the quark ones obey LU, mixed ones are of higher order in ${\cal{L}}_{\rm eff}$.
These operators are constrained by electroweak and top observables, or mixing \cite{Efrati:2015eaa,Brivio:2019ius} and are negligible for the purpose of this work.
Therefore, the (axial-)vector operators \eqref{eq:ops}, which are invariant under QCD-evolution \cite{Alonso:2013hga}, provide a model-independent basis for the
description of dineutrino modes. Effects from
electroweak renormalization group running \cite{Feruglio:2017rjo} can be neglected for  the precision aimed at with this study, since they represent a correction of less than 5\% for $\Lambda_{\text{NP}}\sim 10$\,TeV~\cite{Bause:2021ply}.

\begin{figure}
    \begin{center}
    \includegraphics[width=0.43\textwidth]{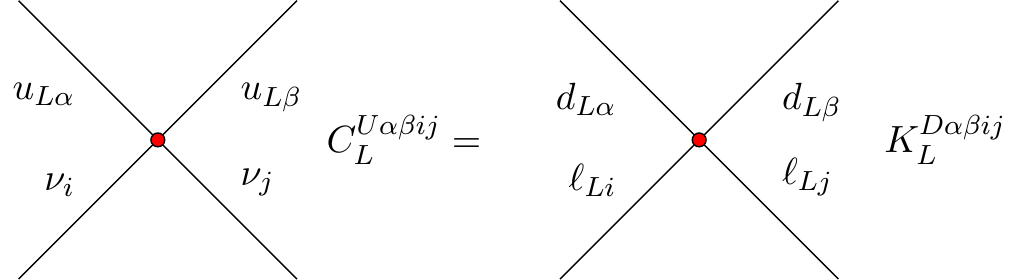}\\
    \hspace{1cm}
    \includegraphics[width=0.43\textwidth]{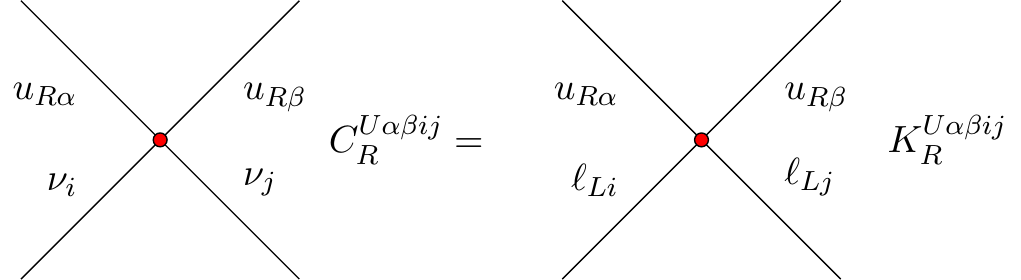}
    \end{center}
    \caption{NP contributions to up and down quark transitions with dineutrinos $C^U_{L,R}$ and dileptons $K^D_{L}$ and $K^U_{R}$, from the operators \eqref{eq:ops}
    in the fermions' flavor basis, with flavor indices $\alpha, \beta,i,j$. The  $SU(2)_L$-based relation between neutrino and charged lepton operators is exploited in \eqref{eq:super} in the mass eigenstate basis.}
    \label{fig:feyn}
\end{figure}

Writing the operators contained in \eqref{eq:ops} into $SU(2)_L$-components one can read off couplings to dineutrinos {$(C_A^M)$} and to charged dileptons {$(K_A^M)$}, where {$M=U$} ({$M=D$}) refers to the up-quark sector (down-quark sector) and $A=L (R)$ denotes left- (right-) handed quark currents
\begin{align} \nonumber
C_L^U&=K_L^D={\frac{2\pi}{\alpha_e}}\left(C^{(1)}_{\ell q} + C^{(3)}_{\ell q}\right)  ,  C_R^U=K_R^U={\frac{2\pi}{\alpha_e}}C_{\ell u}  \,, \\
C_L^D&=K_L^U={\frac{2\pi}{\alpha_e}}\left(C^{(1)}_{\ell q}  - C^{(3)}_{\ell q}\right)  ,
C_R^D=K_R^D={\frac{2\pi}{\alpha_e}}C_{\ell d}  \,, \nonumber
\end{align}
where $\alpha_e$ denotes the fine-structure constant.
The loop factor in the implicit definitions of {$C_A^M,K_A^M$} is included to facilitate matching onto the weak effective theory, $\mathcal{L}=4\, G_F/\sqrt{2}\, (\alpha_e/4 \pi) \sum C_i O_i$.
One observes that\linebreak {$C_R^M=K_R^M$}. While {$C_L^M$} is not fixed in general by {$K_L^M$} due to the different relative signs between $C^{(1)}_{\ell q}$ and $C^{(3)}_{\ell q}$, $C_L^U$ is related to $K_L^D$ and $C_L^D$ to $K_L^U$. 
This link is visualized in Fig.~\ref{fig:feyn}.

Going to mass eigenstates $Q_\alpha=(u_{L \alpha}, V_{\alpha \beta} d_{L \beta})$, $L_i=(\nu_{Li} , W_{ki}^* \ell_{L k})$ with the Cabibbo-Kobayashi-Maskawa \linebreak(CKM)
and Pontecorvo-Maki-Nakagawa-Sakata \linebreak(PMNS) matrices $V$ and $W$, respectively, 
and summing lepton flavors $i,j$ incoherently, we obtain the trace identity
\begin{align}\nonumber
   \sum_{\nu=i,j} & \left( \vert\mathcal{C}_L^{{M} ij}\big\vert^2+\vert\mathcal{C}_R^{{M} ij}\big\vert^2 \right)=\trace \! \left[\mathcal{C}_L^{M}\,\mathcal{C}_L^{{M} \dagger}\,+\,\mathcal{C}_R^{M}\,\mathcal{C}_R^{{M} \dagger}\right]\\      \label{eq:super} 
    = & \trace  \left[ \mathcal{K}_L^{{N}}\mathcal{K}_L^{{N} \dagger}+\mathcal{K}^{M}_R\mathcal{K}_R^{{M} \dagger}\right]  + {\mathcal{O}}(\lambda) \\ 
    =&\sum_{\ell=i,j} \left(   \vert\mathcal{K}_L^{{N} ij}\big\vert^2+\vert\mathcal{K}_R^{{M} ij}\big\vert^2 \right)  + {\mathcal{O}}(\lambda)~,
    \nonumber
\end{align}
between charged lepton couplings $\mathcal{K}_{L,R}$ and neutrino ones $\mathcal{C}_{L,R}$, with Wolfenstein parameter $\lambda\simeq 0.2$. In~\eqref{eq:super}, 
{$M=U$} and {$N=D$} if  the link is exploited for neutrino couplings in the up-quark sector,  as illustrated in Fig.~\ref{fig:feyn}, while 
{$M=D$} and {$N=U$} for dineutrinos coupling to  down-type quarks.
Wilson coefficients in calligraphic style denote those for mass eigenstates. Relation~\eqref{eq:super} follows from {$\mathcal{C}_{L}^{M} = W^\dagger  \,\mathcal{K}_{L}^{N}\, W+ {\mathcal{O}}(\lambda)$}, $\mathcal{C}_{R}^{M}=W^\dagger \,\mathcal{K}_{R}^{M}\, W$  using unitarity.
The traces in \eqref{eq:super} are over the leptonic flavor indices of the Wilson coefficients. In the limit $V=1$, or for $\mathcal{C}_L=\mathcal{K}_L=0$, \eqref{eq:super} becomes an identity. 
The lepton flavor indices $i,j$ are spelled out explicitly in \eqref{eq:super}, while those for the quarks are kept fixed, {{\it e.g.,}} $\alpha=u$, $\beta=c$ {for $c\to u\,\nu\bar\nu$}, and are not given to avoid clutter.
The correction from CKM-rotation for processes involving third and second, (third and first) generations arise at   ${\mathcal{O}}(\lambda^2)$  (at ${\mathcal{O}}(\lambda^3)$)
if  first and second generation mixing can be neglected.

The relation~\eqref{eq:super} allows to predict dineutrino rates for different leptonic flavor structures {$\mathcal{K}_{L,R}^{M \, ij}$} that can be probed with lepton-specific measurements. We identify the following possibilities:
\begin{itemize}
\item[{\it i)}] {$\mathcal{K}_{L,R}^{M \, ij} \propto \delta_{ij}$}, that is, \emph{lepton-universality} (LU).
\item[{\it ii)}] {$\mathcal{K}_{L,R}^{M \, ij}$} are diagonal, that is, \emph{charged lepton flavor conservation} (cLFC).
\item[{\it iii)}] {$\mathcal{K}_{L,R}^{M \, ij}$} general, including \emph{charged lepton flavor violation} (cLFV).
\end{itemize}
Note, { \it ii)} relies on the charged lepton mass basis.
In the following  we discuss implications for  different quark sectors.
We therefore reinstall quark flavor indices and employ the notation
$\mathcal{K}_{L,R}^{tc{ij}}=\mathcal{K}_{L,R}^{U_{23}{ij}}$, ~$\mathcal{K}_{L,R}^{bs{ij}}\,=\mathcal{K}_{L,R}^{D_{23}{ij}}$, $\mathcal{C}_{L,R}^{bsij}=\mathcal{C}_{L,R}^{D_{23}ij} $, and so on.

Limits on the couplings of semileptonic four-fermion operators {$\mathcal{K}_{L,R}^{M \, ij}$} arise from rare decays, and Drell-Yan production.
They are compiled  in Tables~\ref{tab:limits-12s} and \ref{tab:limits-12c} (strange and charm), and Tables~\ref{tab:limits-23b}-\ref{tab:limits-13t} (bottom and top),
with details given in the next sections.

\begin{table*}[h!]
    \centering
    \begin{tabular}{c|cccccc}
    \hline
    \hline
      $sd\ell\ell^\prime$ 
      & $ee$ & $\mu \mu $ & $\tau \tau$  & $e \mu$ & $e \tau$ & $\mu \tau$\\
    \hline
    \hline
  $   | \mathcal{K}_{L,R}^{sd \ell \ell'} |_\text{DY}$ & 
  $3.5$ & $1.9$  & $6.7$ & $2.0$ & $6.1$ & $6.6$ \\
    \hline
     $|  \mathcal{K}_{L,R}^{sd \ell \ell'} |$& ${5\cdot 10^{-2}}$ & ${1.6\cdot 10^{-2}}$  & -- & ${6.6\cdot10^{-4}}$ & -- &  -- \\
    \hline
        $|\mathcal{K}_{R}^{sd \ell \ell'}|_{\nu \bar \nu}^a $ &$[-1.9,0.7]\cdot 10^{-2}$& $[-1.9,0.7]\cdot 10^{-2}$  & $[-1.9,0.7]\cdot 10^{-2}$ & $1.1\cdot 10^{-2}$ & $1.1\cdot 10^{-2}$  & $1.1\cdot 10^{-2}$\\
\hline
    \hline
    \end{tabular}
  \caption{Upper limits on charged dilepton couplings $\mathcal{K}_{A}^{sd\ell\ell^\prime}$ from 
  high--$p_T$ \cite{Fuentes-Martin:2020lea,Angelescu:2020uug} (top row), charged dilepton $K$-decays (mid) and derived ones from $K$ decays to dineutrinos  (bottom). 
  Numbers correspond to a limit on the modulus. 
  LFV-bounds are quoted as flavor-summed, $\sqrt{|\mathcal{K}^{\ell^+ \ell^{\prime-}}|^2 + |\mathcal{K}^{\ell^- \ell^{\prime +}}|^2}$. $^a$Obtained assuming no large cancellations between $\mathcal{K}_{R}^{sd }$ and $\mathcal{K}_{L}^{cu}$.
  }
  \label{tab:limits-12s}
\end{table*}

\section{First and second generation quarks \label{sec:12}}

We exploit the $SU(2)_L$-link between strange and charm.
Constraints on rare $K$ decays are much stronger than those on charm hadrons such that 
the former provides input --obtained in Sec.~\ref{sec:kaons} --  for universality and cLFC tests with $c \to u \nu \bar \nu$ processes, discussed in Sec.~\ref{sec:charm}.

\subsection{Kaon constraints \label{sec:kaons}}

$SU(2)_L$ connects $s\to d\,\nu\bar\nu$ branching ratios with\linebreak charged dilepton couplings $\mathcal{K}^{cu\ell\ell^\prime}_L$ and $\mathcal{K}^{sd\ell\ell^\prime}_R$.
Using \linebreak Ref.~\cite{Mandal:2019gff} and Eq.~\eqref{eq:super}, we obtain
\begin{align}
    &\mathcal{B}(K^+\to \pi^+\,\nu\bar\nu)=A_+^{K^+\pi^+} \cdot x_{sd}^+~,\label{eq:rho_Kp}
\end{align}
where $A_+^{K^+\pi^+}= (68.0 \pm 1.9) \cdot 10^{-8}$~\cite{Brod:2021hsj}, and
\begin{align}
x_{sd}^+=\sum_{i,j}  \big\vert{\mathcal{C}_{\text{SM}}^{sd}} \delta_{ij}+\mathcal{K}_{L}^{{cu} ij} +\mathcal{K}_{R}^{sd ij}\big\vert^2 ~,
\end{align}
with~\cite{Brod:2021hsj}
\begin{align}
\begin{split}
    \mathcal{C}_{\text{SM}}^{sd} = 0.0059-0.0017\,\text{i}\,. \label{eq:CSMsd_val}
\end{split}    
\end{align}
Translating the current measurement \\
$\mathcal{B}(K^+\to\pi^+\nu\bar\nu)_\text{exp} = \left( 8^{+6}_{-4} \right)\cdot 10^{-11}$~\cite{Zyla:2020zbs}
into the $90\%$ CL upper limit $\mathcal{B}(K^+\to\pi^+\nu\bar\nu)_\text{exp} \lesssim 1.7 \cdot 10^{-10}$ yields
\begin{align}
  x_{sd}^+ \lesssim  2.5 \cdot 10^{-4} \,. \label{eq:xpsd_range}
\end{align}
Employing Eqs.~\eqref{eq:CSMsd_val} and \eqref{eq:xpsd_range}, we can extract limits on the couplings $\mathcal{K}_L^{c u}$ and $\mathcal{K}_R^{s d}$.
Since there is only sensitivity to the sum of these couplings, 
we assume that there are no large cancellations and derive bounds only on one coupling at a time.
In the LU limit, we obtain the following limit
\begin{align}
  -0.015 \lesssim \mathcal{K}^{cu\ell\ell}_L,  \mathcal{K}^{sd\ell\ell}_R \lesssim 0.003\,,\quad{\ell=e,\mu,\tau\,.}
\end{align}
Assuming only a single non-vanishing, flavor-conserving contribution, assumed here for concreteness 
to be for ditaus,
the strongest limits corresponding to cLFC read
\begin{align}
    -0.019 \lesssim \mathcal{K}^{cu\tau\tau}_L,  \mathcal{K}^{sd\tau\tau}_R \lesssim 0.007 \,.
\end{align}
Finally, we also consider the case where only one non-diagonal (LFV) coupling is switched on, \textit{e.g.} $\mathcal{K}_{L,R}^{\tau e}=\mathcal{K}_{L,R}^{e\tau}$, which yields
\begin{align}
       |\mathcal{K}^{cu\ell\ell^\prime}_L|,  |\mathcal{K}^{sd\ell\ell^\prime}_R|\lesssim 0.008 \, , \quad \text{for} \quad \ell \neq \ell^\prime \, . 
\end{align}
Comparing these bounds with the corresponding limits from high-$p_T$ in Tables~\ref{tab:limits-12s} and \ref{tab:limits-12c}, we observe that dineutrino data provides stronger constraints, by several orders of magnitude on left-handed $c\to u$ and right-handed  $s\to d$ dilepton couplings. 
Limits from charged dilepton data (mid rows in Tables~\ref{tab:limits-12s} and \ref{tab:limits-12c}) are worked out as well using Refs.~\cite{Mandal:2019gff,Bause:2019vpr,Gisbert:2020vjx}.  

The strongest limits on the real part of $s\,d\,e^+\,e^-$ and $s\,d\,\mu^+\,\mu^-$ couplings are set by $K_L^0\to e^+ e^-$ and $K_L^0\to \mu^+ \mu^-$ decays, while the strongest limit on the imaginary part is provided by $K_L^0\to\pi^0 e^+ e^-$ and $K_L^0\to \pi^0\mu^+ \mu^-$ decays. 
The numbers in Table~\ref{tab:limits-12s} are quoted as the absolute values. 
We also provide limits on $s\,d\,\mu\,e$ couplings which result from $K_L^0\to e^\pm \mu^\mp$. 
In $s\to d$ transitions, dineutrino modes provide the strongest constraints on modes with taus, $\tau^+\tau^-$, $\tau\mu$, and  $\tau e$ which again are kinematically not accessible from kaon decays to charged dileptons. 

\begin{table*}[h!]
    \centering
    \begin{tabular}{c|cccccc}
    \hline
    \hline
      $cu\ell\ell^\prime$ 
      & $ee$ & $\mu \mu $ & $\tau \tau$  & $e \mu$ & $e \tau$ & $\mu \tau$\\
    \hline
    \hline
  $   | \mathcal{K}_{L,R}^{cu \ell \ell'} |_\text{DY}$ & 
  $2.9$ & $1.6$ & $5.6$ & $1.6$ & $4.7$ & $5.1$ \\
    \hline
    $| \mathcal{K}_{L,R}^{cu \ell \ell'} |$ &   $4.0$ & $0.9$ & -- & $2.2$ & n.a.$^\dagger$ & --  \\
    \hline
        $|\mathcal{K}_{L}^{cu \ell \ell'}|_{\nu \bar \nu}^a $ & $[-1.9,0.7]\cdot 10^{-2}$& $[-1.9,0.7]\cdot 10^{-2}$  & $[-1.9,0.7]\cdot 10^{-2}$ & $1.1\cdot 10^{-2}$  & $1.1\cdot 10^{-2}$  & $1.1\cdot 10^{-2}$ \\
\hline
    \hline
    \end{tabular}
  \caption{
  Upper limits on charged dilepton couplings $\mathcal{K}_{A}^{cu\ell\ell^\prime}$ from 
  high--$p_T$ \cite{Fuentes-Martin:2020lea,Angelescu:2020uug} (top row), charged dilepton $D$-decays \cite{Bause:2019vpr,Gisbert:2020vjx} (mid)
  and  derived ones from kaon decays to dineutrinos (bottom).
  Numbers  correspond to a limit on the modulus. 
   LFV-bounds are quoted as flavor-summed, $\sqrt{|\mathcal{K}^{\ell^+ \ell^{\prime-}}|^2 + |\mathcal{K}^{\ell^- \ell^{\prime +}}|^2}$. $^\dagger$ No limit on $D^0 \to e^\pm \tau^\mp$ available. $^a$Obtained assuming no large cancellations between $\mathcal{K}_{R}^{sd }$ and $\mathcal{K}_{L}^{cu}$. }
  \label{tab:limits-12c}
\end{table*}

\subsection{Predictions for charm \label{sec:charm}}

Here we discuss the ramifications of \eqref{eq:super} for $c \to u\, \nu \bar \nu$ processes, which constitute clean null tests of the SM. We show that branching ratios of rare charm hadron decays to dineutrinos cannot exceed upper limits for a given lepton flavor benchmark
{\it i)}-{\it iii)}. 
The situation  for $c \to u \,\nu \bar \nu$  dineutrino transitions is quite unique as the SM amplitude is entirely negligible due to an efficient  GIM-cancellation \cite{Burdman:2001tf} and the current lack of experimental constraints~\footnote{The $D^0 \to \nu \bar \nu$ branching ratio induced by (axial-)vector operators is helicity suppressed by two powers of neutrino mass, and negligible. The Belle result ${\mathcal{B}}(D^0 \to \nu \bar \nu) < 9.4 \cdot 10^{-5}$ at 90$\,\%$ CL~\cite{Lai:2016uvj} can therefore be safely avoided.}. 
Further details can be found in Ref.~\cite{Bause:2020xzj}.

Since the neutrino flavors are not tagged, as common to generic particle physics experiments,  
a dineutrino branching ratio is an incoherent sum of flavor-specific ones
\begin{align} \label{eq:xcu}
&\mathcal{B}\left( c \to u \,\nu \bar \nu\right)=\sum_{i,j} \mathcal{B}\left( c\to u \,\nu_i  \bar \nu_j\right) \propto x_{cu}\, , \\
&x_{cu}=\sum_{i,j}|\mathcal C_L^{U ij}|^2+ |\mathcal C_R^{U ij}|^2 \, .  \nonumber 
\end{align}
Using \eqref{eq:super} with $M=U$ and $N=D$ and 
the constraints on $|\Delta s|=|\Delta d|=1$ and $|\Delta c|=|\Delta u|=1$  couplings given in Tables  \ref{tab:limits-12s}-\ref{tab:limits-12c} 
we obtain upper limits in  the lepton flavor benchmarks {\it i)}-{\it iii)}:
\begin{align}  \label{eq:LU}
x_{cu} &= 3\, x_{cu}^{\mu \mu} \lesssim \,2.6 \,, \quad  (\text{LU}) \\ \label{eq:cLFC}
x_{cu} &= x_{cu}^{e e}\hspace{-0.1cm}+ x_{cu}^{\mu \mu }\hspace{-0.1cm}+x_{cu}^{\tau \tau}  \lesssim 156 \,, \quad (\text{cLFC}) \\ \label{eq:total}
x_{cu} &= x^{ee} \hspace{-0.1cm}+ x_{cu}^{\mu \mu}\hspace{-0.1cm} +x_{cu}^{\tau \tau} \hspace{-0.1cm}+2 ( x_{cu}^{e \mu}+ x_{cu}^{e \tau}+x_{cu}^{ \mu \tau} ) \lesssim 655  \,.
\end{align}
Here, the flavor-specific contributions $x_{cu}^{\ell \ell^\prime}= | \mathcal{K}_{L}^{{sd} \ell \ell^\prime}|^2+ |\mathcal{K}_{R}^{cu \ell \ell^\prime}|^2 +{\cal{O}}(\lambda)$
, where $x_{cu}=\sum_{\ell,\ell^\prime} x_{cu}^{\ell \ell^\prime}$, to each upper limit have been spelled out.
Since dimuon bounds are the most stringent ones  they govern the LU-limit \eqref{eq:LU}. 
Note also that subleading CKM-corrections are included in \eqref{eq:LU}-\eqref{eq:total}~\cite{Bause:2020xzj}.

In contrast to Ref.~\cite{Bause:2020xzj} we use low energy constraints on the couplings $\mathcal{K}_{A}^{qq^\prime\ell{\ell^\prime}}$ in addition to the high-$p_T$ ones, \textit{i.e.} we use the strongest available limits on $\mathcal{K}_{L,\,R}^{sd\ell{\ell^\prime}}$ and $\mathcal{K}_{L,\,R}^{cu\ell{\ell^\prime}}$ from Tables~\ref{tab:limits-12s} and~\ref{tab:limits-12c}. The inclusion of rare kaon data implies improved bounds in Eqs.~\eqref{eq:LU}, \eqref{eq:cLFC}, \eqref{eq:total} and Table~\ref{tab:afactors} with respect to the results in Ref.~\cite{Bause:2020xzj}.

Various exclusive dineutrino branching ratios depend on combinations of left and right quark chiralities, 
\begin{align}\label{eq:xpmgen-c}
       x_{cu}^\pm =\sum_{i,j}|\mathcal C_L^{cu ij} \pm \mathcal C_R^{cu ij}|^2
\end{align}
with $x_{cu}^+ + x_{cu}^-=2 x_{cu}$, hence both $x_{cu}^\pm$ are bounded by $ x_{cu}$.
Specifically, $x^+_{cu}$ arises in the $D \to P \,\nu \bar \nu$ branching ratio, whereas  $D \to P \,P'\, \nu \bar \nu$, $P,P'=\pi,K$, branching ratios are dominated by 
$x^-_{cu}$. Inclusive decays and, approximately, the baryonic ones are dominated by $x_{cu}$.
All branching ratios  of a charmed hadron $h_c$ into a final hadron $F$ and dineutrinos can be  written as
\begin{align}\label{eq:BR-A}
    \mathcal{B}(h_c\to F\,\nu \bar \nu)= A_+^{h_c F} \, x^+_{cu} + A_-^{h_c F} \, x^-_{cu} \,,
\end{align}
with  $A_\pm^{h_c F}$ coefficients \cite{Bause:2020xzj} compiled  in Table~\ref{tab:afactors}.
Experimental extraction of $x_{cu}$ or $x^\pm_{cu} \leq 2 \,x_{cu}$ above the upper limit in \eqref{eq:LU} would indicate a breakdown of LU,
whereas values above \eqref{eq:cLFC} would imply cLFV. 
We also obtain achievable, total upper limits in \eqref{eq:total}. 
Corresponding upper limits on branching ratios of rare charm dineutrino modes are given in Table~\ref{tab:afactors}. 

We recall that $c \to u$ dineutrino FCNCs require the presence of NP to be observable. Due to their strong ties with the charged dilepton modes, dineutrino modes are not only clean probes of BSM physics, their rate  is informative on lepton flavor.
\begin{table}[ht!]
 \centering
    \resizebox{0.45\textwidth}{!}{
  \begin{tabular}{lccccc}
  \hline
  \hline
  $h_c\to F $ & $A^{h_c\,F}_+$ & $A^{h_c\,F}_-$ &  $\mathcal{B}_{\text{LU}}^\text{max}$ & $\mathcal{B}_{\text{cLFC}}^\text{max}$ & $\mathcal{B}^\text{max}$ \\
  &$[10^{-8}]$&$[10^{-8}]$& $[10^{-7}]$ &$[10^{-6}]$ & $[10^{-6}]$\\
  \hline
  $D^0\to\pi^0$ & $0.9$ & -- & $0.5$ & $2.8$ & $ 12$ \\
  $D^+\to\pi^+$ & $3.6$ & -- & $1.9$ & $11$ & $47$ \\
  $D_s^+\to K^+$ & $0.7$ & --  & $0.3$ & $2.1$ & $8.8$ \\
  &&&&   & \\
  $D^0\to\pi^0\pi^0$ & $\mathcal{O}(10^{-3})$ & $0.21$ & $ 0.1$ & $0.7$ & $2.8$ \\
  $D^0\to\pi^+\pi^-$ & $\mathcal{O}(10^{-3})$ & $0.41$ & $ 0.2$ & $1.3$ & $5.4$ \\
  $D^0\to K^+K^-$ & $\mathcal{O}(10^{-6})$ &  $0.004$ & $0.002$ & $0.01$ & $0.06$ \\
  &&&& & \\
  $\Lambda_c^+\to p^+$ & $1.0$ & $1.7$ & $1.4$ & $8.4$ & $35$ \\
  $\Xi_c^+\to \Sigma^+$ & $1.8$ & $3.5$ & $2.7$ & $ 17$ & $70$ \\ 
  &&&&   &\\
  $D^0\to X$ & $2.2$ & $2.2$ & $1.1$ & $6.9$ & $29$ \\
  $D^+\to X$ & $5.6$ & $5.6$ & $2.9$ & $17$ & $74$ \\
  $D_s^+\to X$ & $2.7$ & $2.7$ & $1.4$ & $8.3$ & $35$ \\
  \hline
  \hline
   \end{tabular}}
\caption{Coefficients $A^{h_c\,F}_\pm$, as defined in \eqref{eq:BR-A}, and model-independent upper limits on $\mathcal{B}_{\text{LU}}^{\text{max}}$, $\mathcal{B}_{\text{cLFC}}^{\text{max}}$, $\mathcal{B}^{\text{max}}$ from \eqref{eq:LU}, \eqref{eq:cLFC} and \eqref{eq:total}, respectively, corresponding to  the lepton flavor symmetry benchmarks {\it i)}-{\it iii)}. The first two columns are taken from Ref.~\cite{Bause:2020xzj}.
}
\label{tab:afactors}
\end{table}

\begin{table}[h!]
    \centering
    \resizebox{0.45\textwidth}{!}{
    \begin{tabular}{c|cccccc}
    \hline
    \hline
      $bs\ell\ell^\prime$ 
      & $ee$ & $\mu \mu $ & $\tau \tau$  & $e \mu$ & $e \tau$ & $\mu \tau$\\
    \hline
    \hline
 $   | \mathcal{K}_{L,R}^{bs \ell \ell'} |_\text{DY}$ & $13$ & $7.1$ & $25$ & $8.0$ & $27$ & $30$ \\
    \hline
    $ \mathcal{K}_{R}^{bs \ell \ell'} $ & $0.04$ & $[-0.03; -0.01]$ & $32$ & $0.1$ & $2.8$ & $3.4$ \\
      $\mathcal{K}_{L}^{bs \ell \ell'} $& $0.04$ & $[-0.06; -0.04]$ & $32$ & $0.1$ & $2.8$ & $3.4$ \\
    \hline
    $\mathcal{K}_{R}^{bs \ell \ell'}|_{\nu \bar \nu} $ & $1.4$ & $1.4$ & $1.4$ & $1.8$ & $1.8$ & $1.8$ \\
\hline
    \hline
    \end{tabular}}
  \caption{
  Upper limits on charged dilepton couplings $\mathcal{K}_{A}^{bs\ell\ell^\prime}$ from 
  high--$p_T$ \cite{Fuentes-Martin:2020lea,Angelescu:2020uug} (top row), charged dilepton $B$-decays (middle rows) and derived ones   from three-body rare $B$-decays to dineutrinos (bottom). Numbers without ranges correspond to a limit on the modulus. the $\mu \mu$ ranges are obtained from a global fit, with the departures from zero in $\mathcal{K}_{L}^{bs \mu \mu} $ corresponding to the flavor anomaly.
   LFV-bounds are quoted as flavor-summed, $\sqrt{|\mathcal{K}^{\ell^+ \ell^{\prime-}}|^2 + |\mathcal{K}^{\ell^- \ell^{\prime +}}|^2}$, whereas the other bounds are for a single coupling.
  }
  \label{tab:limits-23b}
\end{table}

\begin{table}[h!]
    \centering
    \resizebox{0.45\textwidth}{!}{    
    \begin{tabular}{c|cccccc}
    \hline
    \hline
      $bd\ell\ell^\prime$ 
      & $ee$ & $\mu \mu $ & $\tau \tau$  & $e \mu$ & $e \tau$ & $\mu \tau$\\
    \hline
    \hline
  $| \mathcal{K}_{L,R}^{bd \ell \ell'} |_\text{DY}$ & 
    $5.0$ & $2.7$ & $9.6$ & $3.1$ & $9.6$ & $11$ \\
    \hline
    $ \mathcal{K}_{R}^{bd \ell \ell'} $ &  $0.09$ & $[-0.03,0.03]$ & $21$ & $0.2$ & $3.4$ & $2.4$ \\
    $ \mathcal{K}_{L}^{bd \ell \ell'} $ &  $0.09$ & $[-0.07,0.02]$ & $21$ & $0.2$ & $3.4$ & $2.4$ \\
    \hline
    $\mathcal{K}_{R}^{bd \ell \ell'}|_{\nu \bar \nu} $ & $1.8$ & $1.8$ & $1.8$ & $2.5$ & $2.5$ & $2.5$ \\
\hline
    \hline
    \end{tabular}
    }
  \caption{
  Upper limits on charged dilepton couplings $\mathcal{K}_{A}^{bd\ell\ell^\prime}$ from 
  high--$p_T$ \cite{Fuentes-Martin:2020lea,Angelescu:2020uug} (top row), charged dilepton $B$-decays (middle rows) and
  derived ones from three-body rare $B$-decays to dineutrinos  (bottom). Numbers without ranges correspond to a limit on the modulus. The $\mu \mu$ ranges are
  obtained from a global fit.
   LFV-bounds are quoted as flavor-summed, $\sqrt{|\mathcal{K}^{\ell^+ \ell^{\prime-}}|^2 + |\mathcal{K}^{\ell^- \ell^{\prime +}}|^2}$, whereas the other bounds are for a single coupling. 
  }
  \label{tab:limits-13b}
\end{table}

\section{Third generation quarks  \label{sec:3}}

We summarize predictions (Sec.~\ref{sec:beauty}) and  tests  (Sec.~\ref{sec:LU-beauty})
for beauty decays
and work out  predictions and opportunities for the top sector in Sec.~\ref{sec:top} and \ref{sec:1top}.

\subsection{Predictions for beauty \label{sec:beauty}}

In this section we study $b\to q\,\nu\bar\nu$ transitions and their interplay with $b\to q\,\ell^+\ell^-$ transitions routed by \eqref{eq:super}. 
First, we use SM effective theory to improve the limits on charged ditau couplings from dineutrino data. 
Finally, we exploit Eq.~\eqref{eq:super} using the complementarity between $B\to V$ and $B\to P$ dineutrino branching ratios, which offers novel tests of lepton universality. 
Further details can be found in Ref.~\cite{Bause:2021ply}. 

Using the current experimental $90\,\%$ CL upper limits~\cite{Zyla:2020zbs}
\begin{align} 
\begin{split}
\mathcal{B}(B^0 \to K^{*0}\nu\bar{\nu})_\text{exp}&< 18 \cdot 10^{-6} \,, \\
\mathcal{B}(B^+ \to K^+\nu\bar{\nu})_\text{exp}& < 16 \cdot 10^{-6} \,, 
\label{eq:bnunuexp}
\end{split} \\
\begin{split}
\mathcal{B}(B^+ \to \rho^+\nu\bar{\nu})_\text{exp}&< 30\cdot 10^{-6} \, , \\
\mathcal{B}(B^+\to \pi^+\nu\bar{\nu})_\text{exp}&< 14\cdot 10^{-6} \,,    
\end{split}
\end{align}
one can extract the following bounds on $x_{bq}^{\pm}$~\cite{Bause:2021ply}
\begin{align}
    &x_{bs}^{+}\lesssim 2.9,\qquad x_{bs}^- +0.2\, x_{bs}^+ \lesssim 2.0~,\label{eq:bound1}\\
    &x_{bd}^{+}\lesssim 4.2,\qquad x_{bd}^- +0.1 \,x_{bd}^+ \lesssim 2.4~,\label{eq:bound2}
\end{align}
where
\begin{align}\label{eq:xpmgen}
\begin{split}
       x_{bs}^\pm = \sum_{i,j}  \big\vert{\mathcal{C}_{\text{SM}}^{bs}}\,\delta_{ij}+\mathcal{K}_{L}^{{tc} ij}\pm\mathcal{K}_{R}^{bs ij}\big\vert^2 ~,\\
       x_{bd}^\pm = \sum_{i,j}  \big\vert {\mathcal{C}_{\text{SM}}^{bd}}\,\delta_{ij}+\mathcal{K}_{L}^{{tu} ij}\pm\mathcal{K}_{R}^{bd ij}\big\vert^2 ~.
\end{split}
\end{align}
These are the analogues of \eqref{eq:xpmgen-c} for beauty and include the finite SM contribution
$\mathcal{C}_{\text{SM}}^{bs} = 0.50-0.01\,\text{i}$ and $\mathcal{C}_{\text{SM}}^{bd} = -0.10-0.04\,\text{i}$.

The limits \eqref{eq:bound1} and \eqref{eq:bound2} allow one  to set bounds on $\mathcal{K}_{L}^{{tc(u)} ij}$, $\mathcal{K}_{R}^{bs(d) ij}$ via Eq.~\eqref{eq:xpmgen}, depending on the lepton flavor assumptions. We observe that limits from charged lepton data are stronger or similar than the dineutrino bounds with the exception of $\tau\tau$ and $e \tau$ for $b\to d$,~\cite{Bause:2021ply} 
\begin{align}
    |\mathcal{K}_{R}^{bd\tau\tau}|\lesssim 1.8~,\quad   |\mathcal{K}_{R}^{bde\tau}|\lesssim 2.5~, 
\end{align}
and $e\tau$, $\mu\tau$, $\tau\tau$  for $b\to s$
\begin{align}
    |\mathcal{K}_{R}^{bs\tau\ell}|\lesssim 1.4~,\qquad|\mathcal{K}_{R}^{bs\tau\tau}|\lesssim 1.8\,.
\end{align}
Notice that dineutrino limits are stronger than those from Drell-Yan data, cf. Tables~\ref{tab:limits-23b} and \ref{tab:limits-13b}.
The constraints on the top FCNCs following from the same analysis are given in Tables~\ref{tab:limits-23t} and \ref{tab:limits-13t}.

\subsection{Testing universality with $b\to q\,\nu\bar\nu$ \label{sec:LU-beauty}}

\begin{figure}[ht!]
    \centering
    \includegraphics[scale=0.53]{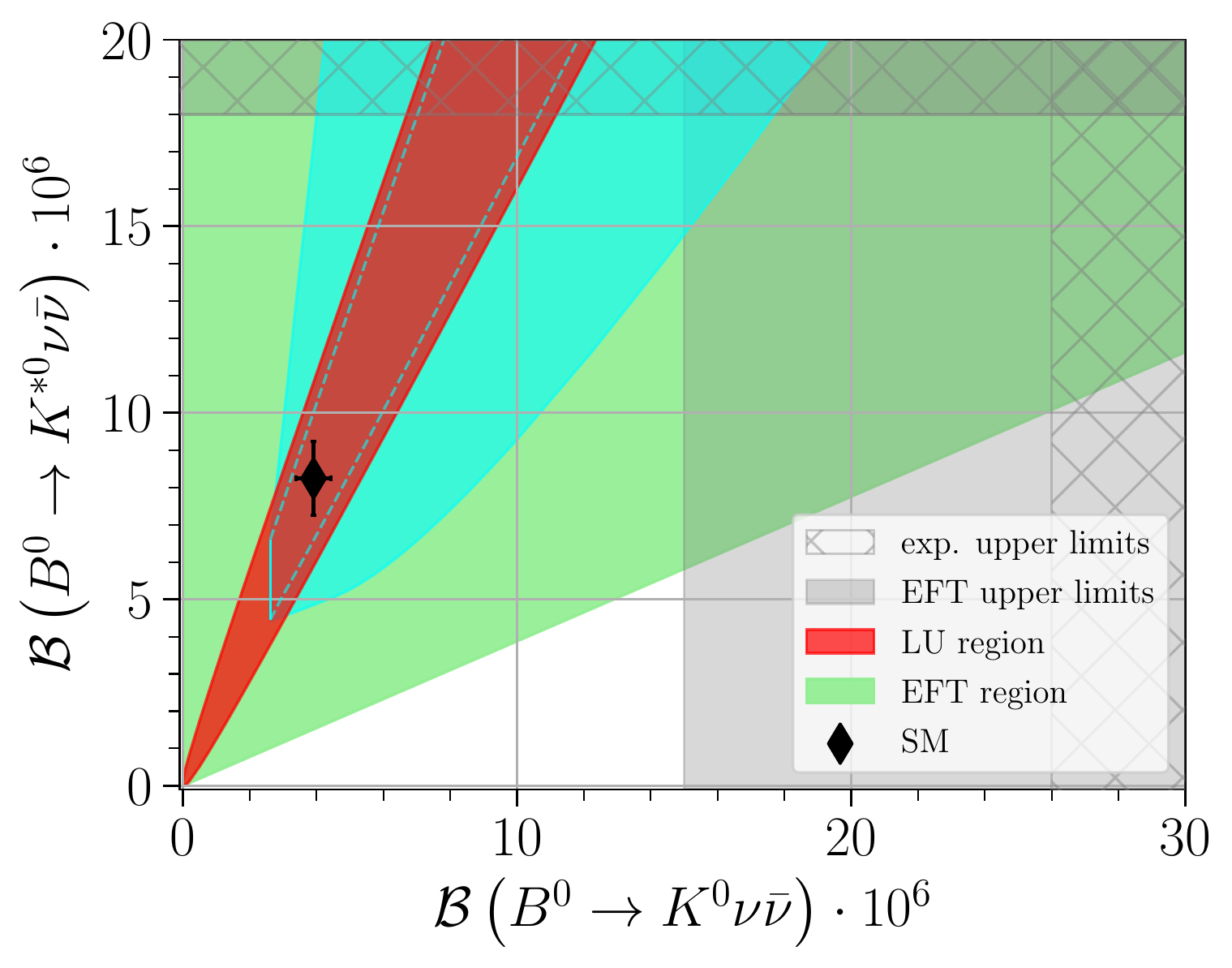}
   \caption{$\mathcal{B}(B^0 \to K^{*0}\,\nu\bar{\nu})$ versus $\mathcal{B}(B^0 \to K^0\,\nu\bar{\nu})$.
    SM predictions with uncertainties (black) from \eqref{eq:bnunuSM}. The red region represents the LU region \eqref{eq:luregion} at $1\,\sigma$, and it is mainly dominated from form factor uncertainties. The green region represents the validity of the EFT framework, given by Eq.~\eqref{eq:bound1}. Cyan region represents cLFC region for a single species, in particular $\ell=\tau$, assuming a bound on $\mathcal{B}(B_s^0\to\tau^+\tau^-)\lesssim 10^{-6}$. Dashed cyan lines represent the cLFC region for $\ell=\mu$, see main text.
   }
    \label{fig:plotKstarversusK}
\end{figure}

The branching ratios for $B\to V\,\nu\bar\nu$ and $B\to P\,\nu\bar\nu$ decays in the lepton universality limit are given by
\begin{align}\label{eq:LUBtoV}
    &\mathcal{B}(B\to V\,\nu\bar\nu)_{\text{LU}}\,=\,A_+^{BV}\,x_{bq,\text{LU}}^+ +\,A_-^{BV}\,x_{bq,\text{LU}}^-~,\\
    &\mathcal{B}(B\to P\,\nu\bar\nu)_{\text{LU}}\,=\,A_+^{BP}\,x_{bq,\text{LU}}^+~, \label{eq:LUBtoP}
\end{align}
respectively, with $x_{bq,\text{LU}}^\pm\,=\,3\,\left|\mathcal{C}_{\text{SM}}^{bq\ell\ell}+\mathcal{K}_{L}^{tq^\prime\ell\ell}\pm\mathcal{K}_{R}^{bq\ell\ell}\right|^2$~, with $q^\prime=u,\,(c)$ for $q=d,\,(s)$, respectively, and  $\ell$ fixed to the flavor  with strongest constraints.
Since present rare top data are not able to put useful constraints on  the coupling $\mathcal{K}_{L}^{tq^\prime\ell\ell}$,
we instead solve $\mathcal{B}(B\to P\,\nu\bar\nu)_{\text{LU}}$ in Eq.~\eqref{eq:LUBtoP}  for  $\mathcal{K}_{L}^{tq^\prime\ell\ell}$ 
and plug the two solutions into Eq.~\eqref{eq:LUBtoV}.
This yields a correlation between the  branching ratios,
\begin{align}\label{eq:luregion}
    &\mathcal{B}(B\to V\,\nu\bar\nu)_{\text{LU}}\,=\,\frac{A_+^{BV}}{A_+^{BP}}\,\mathcal{B}(B\to P\,\nu\bar\nu)_{\text{LU}}\\
    &\,+\,3\,A_{-}^{BV}\,\left|\sqrt{ \frac{\mathcal{B}(B\to P\,\nu\bar\nu)_{\text{LU}}}{3\, A_+^{BP} }}\mp 2\,\mathcal{K}_{R}^{bq\ell\ell}\right|^2~.\nonumber
\end{align}
The most stringent limits on $\mathcal{K}_{R}^{bq\ell\ell}$ are given for muons. By performing a six-dimensional global fit of the semi-leptonic Wilson coefficients $\mathcal{C}_{(7,9,10),\mu}^{(\prime)}$ to current experimental information on $b\to s\,\mu^+\mu^-$ data (excluding $R_{K^{(*)}}$ which can be polluted by NP effects in electron couplings), we obtain the following $1\sigma$ fit values~\cite{Bause:2021ply,Bause:inprep}
\begin{align}
  \mathcal{K}_{R}^{bs\mu\mu}&=V_{tb}V^{\ast}_{ts}\,(0.46\pm 0.26)\,,\label{eq:KRglobalfit1sigma_bs}\\
\mathcal{K}_{R}^{bd\mu\mu}&=V_{tb}V^{\ast}_{td}\,(0\pm 4)\,.\label{eq:KRglobalfit1sigma_bd}
\end{align}
Fig.~\ref{fig:plotKstarversusK} displays the correlation between $\mathcal{B}(B^0 \to K^{*0}\nu\bar{\nu})$ and $\mathcal{B}(B^0 \to K^0\nu\bar{\nu})$ using Eq.~\eqref{eq:luregion}, with $\mathcal{K}_{R}^{bs\mu\mu}$ from Eq.~\eqref{eq:KRglobalfit1sigma_bs}. 
Scanning $\mathcal{K}_{R}^{bs\mu\mu}$, and the form factors~\cite{Bause:2021ply}
$A_+^{B^0 K^{*0}}=(200 \pm 29) \cdot 10^{-8}$,  $A_-^{B^0 K^{*0}}=(888 \pm 108) \cdot 10^{-8}$, $A_+^{B^0 K^{0}}=(516 \pm 68) \cdot 10^{-8}$, within their $1\,\sigma$ ($2\,\sigma$) regions,
one obtains 
\begin{align}
\frac{\mathcal{B}(B^0 \to K^{*0}\nu\bar{\nu})_{\text{LU}}}{\mathcal{B}(B^0 \to K^{0}\nu\bar{\nu})_{\text{LU}}}=1.7\ldots2.6 \quad (1.3\ldots 2.9)\, ,
\end{align}
displayed by the red region.
Measurements outside this region would signal a breakdown of LU. 
Note, however, that  a measurement inside this region does not necessarily imply LU. 
The SM predictions 
\begin{align} 
\begin{split}
\mathcal{B}(B^0 \to K^{*0}\nu\bar{\nu})_\text{SM}&=(8.2\,\pm\,1.0)\cdot 10^{-6} \, , \\
\mathcal{B}(B^0\to K^0\nu\bar{\nu})_\text{SM}&=(3.9\,\pm\,0.5)\cdot 10^{-6} \,,
\label{eq:bnunuSM}
\end{split}
\end{align}
are depicted with their uncertainties (black). The green region represents the validity of our effective field theory (EFT) framework, given by Eq.~\eqref{eq:bound1}. 
The hatched bands correspond to current experimental $90\,\%$ CL upper limits  (\ref{eq:bnunuexp}). The gray bands represent the derived EFT limits, $\mathcal{B}(B^0 \to K^0\nu\bar{\nu})_{\text{derived}}\,<\,1.5\,\cdot\, 10^{-5}$~\cite{Bause:2021ply}. 
A measurement between gray and hatched area would imply BSM physics not covered by the EFT framework. 
Belle II is expected to observe these modes with 50 ab$^{-1}$ and $10 \%$ accuracy.
Similar conclusions are found in $b\to d\,\nu\bar\nu$ decay, a detailed analysis can be found in Ref.~\cite{Bause:2021ply}.

The possibility of testing cLFC mainly depends on \linebreak $\mathcal{K}_{R}^{bs\tau\tau}$ constraints. Currently, the bounds provided by dineutrino modes are a factor 23 stronger than direct limits from $\mathcal{B}(B_s^0\to\tau^+\tau^-)$, see Table~\ref{tab:limits-23b}. Working out the projections from Belle II with $5~$ab$^{-1}$, $\mathcal{B}(B_s^0\to\tau^+\tau^-)_{\text{proj}}\leq 8.1\cdot 10^{-5}$~\cite{Kou:2018nap}, we observe that dineutrino limits would still be a factor 3 stronger. In contrast if the direct experimental limit would be two orders of magnitude stronger $\mathcal{B}(B_s^0\to\tau^+\tau^-)\lesssim 10^{-6}$, testing cLFC would be possible. Figure~\ref{fig:plotKstarversusK} displays precisely the cLFC limit for single species ($\ell$ fixed), that is $x^\pm_{bs}\,=\,2\,\big\vert\mathcal{C}_{\text{SM}}^{bs}\big\vert^2+\big\vert{\mathcal{C}_{\text{SM}}^{bs}}+\mathcal{K}_{L}^{{tc} \ell\ell}\pm\mathcal{K}_{R}^{bs \ell\ell}\big\vert^2$. For $\ell=\tau$ (filled cyan region), we assume $\mathcal{B}(B_s^0\to\tau^+\tau^-)\lesssim 10^{-6}$, which implies a constraint on $\mathcal{K}_{R}^{bs \tau\tau}$, while $\mathcal{K}_{L}^{{tc} \tau\tau}$ remains unconstrained. The filled cyan region is then obtained by varying $\mathcal{K}_{R}^{bs \tau\tau}$ and $\mathcal{K}_{L}^{{tc} \tau\tau}$. A measurement outside of this region implies a violation of cLFC in the limit of large $\tau\tau$ couplings. We have also studied $\ell=\mu$ indicated by the dashed cyan lines, where the corresponding region lies inside the LU region, because $\mathcal{K}_{R}^{bs \mu\mu}$ is strongly bounded from $b\to s\mu\mu$ global fits.
Note that assuming cLFC for a single species, the region is bounded from below as $x_{bs}^\pm \geq 2\,|\mathcal{C}_\text{SM}^{bs}|^2$ (cyan solid line), implying a lower limit on the branching ratios of $2/3$ of the SM one.

\subsection{Predictions for top \label{sec:top}}

The $SU(2)_L$--link connects  left-handed quark couplings in $b\to s\ell\ell^{(\prime)}$ and $b\to d\ell\ell^{(\prime)}$ observables to 
left-handed $tc$-- and $tu$--dineutrino couplings, respectively. \linebreak Constraints on the former, $\mathcal{K}_L^{bsij}$ and $\mathcal{K}_L^{bdij}$,
are available for all lepton flavors $i,j$, see Tables~\ref{tab:limits-23b}-\ref{tab:limits-13b}.
Constraints on right-handed contributions $\mathcal{K}_R^{tqij}$ are much weaker.
The most stringent ones available are on cLFV couplings.
CMS  \cite{CMS:2021nlh} provides upper limits on  $e \mu$ final states,  corresponding to
rare top branching ratios 
\begin{align}   \nonumber
{\cal{B}}(t \to u\, e^+ \mu^- + u\, e^- \mu^+)_{\text{exp}}&<0.135  \cdot 10^{-6} \, , \\
{\cal{B}}(t \to c\, e^+ \mu^- + c\, e^- \mu^+)_{\text{exp}}&<1.31  \cdot 10^{-6} \, . \label{eq:CMS}
\end{align}
ATLAS \cite{ATLAS:2018avw} also provides  
bounds on cLFV modes involving taus (at \@ 95 \% CL)
\begin{align} \label{eq:ATLAS}
\mathcal{B}(t \to q \,\tau \,(e,\mu))_\text{exp} < 1.86 \cdot 10^{-5} \,,  \quad q=u,c  \,, 
\end{align}
 which we scaled to arrive at the constraints on the 
Wilson coefficients in Tables~\ref{tab:limits-23t}-\ref{tab:limits-13t}.
We also employ limits on an admixture of $ tt ee$ and   $ tt \mu \mu$ couplings  \cite{CMS:2020lrr}
as upper limits on the FCNC ones $ tq\ell \ell$, $|\mathcal{K}_{L,R}^{tq\ell \ell}| \lesssim |\mathcal{K}_{L,R}^{tt\ell \ell}| \lesssim 200$ for $\ell=e,\mu$.
This is consistent with the rare mode being considered as a signal in the search.
Projected limits for  140 $\rm{fb}^{-1}$ of LHC data in single top production in association with  dimuons  \cite{Afik:2021jjh} would yield $|\mathcal{K}_{L,R}^{tc\mu \mu}| \lesssim 40$
and $|\mathcal{K}_{L,R}^{tu\mu \mu}| \lesssim 4.8$. These are stronger than the $t \bar t \mu \mu$ constraints, and those from LFV $t \to u$ decays \eqref{eq:CMS},\eqref{eq:ATLAS},
but slightly weaker than the ones from LFV $t \to c$ decays. The constraints on the left-handed coefficients from  rare $B$-decays to  dineutrinos, denoted in 
Tables~\ref{tab:limits-23t} and \ref{tab:limits-13t} by
$| \mathcal{K}_{L}^{tq \ell \ell'} |_{\nu \bar \nu}$,  are the strongest.
Note that all chiralities contribute decoherently in the high energy limit where masses are neglected.
We are not aware of any data  with ditaus.

By means of \eqref{eq:super}, $\mathcal{K}_L^{bqij}$ and $\mathcal{K}_R^{tqij}$
enter dineutrino observables in the top-sector
\begin{align}
 x_{tc}=\sum_{i,\,j} \vert\mathcal{K}_L^{bsij}\vert^2 +\vert\mathcal{K}_R^{tcij}\vert^2  +\mathcal{O}(\lambda) \,, \\
 x_{tu}=\sum_{i,\,j} \vert\mathcal{K}_L^{bdij}\vert^2 +\vert\mathcal{K}_R^{tuij}\vert^2 +\mathcal{O}(\lambda) \,,
\end{align}
where $x_{tc}$ and $x_{tu}$ are defined as in charm  \eqref{eq:xcu}.
The absence of a  limit on $\mathcal{K}_R^{tq\tau \tau}$ allows only to test LU.
We obtain
\begin{align} \label{eq:tq}
x_{tq} \lesssim 1.2 \cdot 10^5\quad\text{(LU)} \, , \quad q=u,c \, .
\end{align}
Significantly stronger limits, and further predictions are obtained when assuming contributions from left-handed couplings only.
We find for $tc\nu\nu$
\begin{equation} \label{eq:tc}
    \begin{split}
    x_{tc}^L&\lesssim 0.011\quad\text{(LU)}\,,\\
    x_{tc}^L&\lesssim 630 \quad\text{(cLFC)}\,,\\
    x_{tc}^L&\lesssim 650 \quad\text{(general)}\,,
    \end{split}
\end{equation}
and for $tu\nu\nu$
\begin{equation}  \label{eq:tu}
    \begin{split}
    x_{tu}^L&\lesssim 0.015 \quad\text{(LU)}\,,\\
    x_{tu}^L&\lesssim 90 \quad\text{(cLFC)}\,,\\
    x_{tu}^L&\lesssim 120\quad\text{(general)}\,.
    \end{split}
\end{equation}
These are the first bounds of this type constraining $tc$-- and $tu$--dineutrino couplings.

\begin{table}[h!]
    \centering
    \resizebox{0.45\textwidth}{!}{
    \begin{tabular}{c|cccccc}
    \hline
    \hline
      $tc\ell\ell^\prime$ 
      & $ee$ & $\mu \mu $ & $\tau \tau$  & $e \mu$ & $e \tau$ & $\mu \tau$\\
    \hline
    \hline
           $| \mathcal{K}_{L,R}^{tc \ell \ell'} |$ &   $\sim 200 $ & $\sim 200$ & n.a. & $36$ & 136 &  136 \\
           \hline
    $| \mathcal{K}_{L}^{tc \ell \ell'} |_{\nu \bar \nu}$ &   ${[-1.9,0.9]}$ &  ${[-1.9,0.9]}$&  ${[-1.9,0.9]}$ & $1.8$ & $1.8$ & $1.8$  \\
\hline
    \hline
    \end{tabular}}
  \caption{
  Upper limits  on charged dilepton couplings $\mathcal{K}_{A}^{tc\ell\ell^\prime}$ from
  collider studies  \cite{CMS:2021nlh,ATLAS:2018avw,CMS:2020lrr} of  top plus charged dilepton processes  (top row), see text,
  and 
    on charged dilepton couplings $\mathcal{K}_{L}^{tc\ell\ell^\prime}$ derived from 
three-body  rare $B$-decays  to dineutrinos (bottom row) except when a range is given. Numbers  correspond to a limit on the modulus. 
   LFV-bounds are quoted as flavor-summed, $\sqrt{|\mathcal{K}^{\ell^+ \ell^{\prime-}}|^2 + |\mathcal{K}^{\ell^- \ell^{\prime +}}|^2}$. 
  }
  \label{tab:limits-23t}
\end{table}

\begin{table}[h!]
    \centering
    \resizebox{0.45\textwidth}{!}{
    \begin{tabular}{c|cccccc}
    \hline
    \hline
      $tu\ell\ell^\prime$ 
      & $ee$ & $\mu \mu $ & $\tau \tau$  & $e \mu$ & $e \tau$ & $\mu \tau$\\
    \hline
    \hline
               $| \mathcal{K}_{L,R}^{tu \ell \ell'} |$ &   $\sim 200$ & $\sim 200$ & n.a. & $12$ & 136 &  136 \\
           \hline
    $| \mathcal{K}_{L}^{tu \ell \ell'} |_{\nu \bar \nu}$ &   ${[-1.6,1.8]}$ & ${[-1.6,1.8]}$ & ${[-1.6,1.8]}$ & ${2.4}$ & ${2.4}$  & ${2.4}$  \\
\hline
    \hline
    \end{tabular}}
  \caption{
     Upper limits  on charged dilepton couplings $\mathcal{K}_{A}^{tu\ell\ell^\prime}$ from
  collider studies  \cite{CMS:2021nlh,ATLAS:2018avw,CMS:2020lrr}  of  top plus charged dilepton processes  (top row), see text, 
  and on charged dilepton couplings $\mathcal{K}_{L}^{tu\ell\ell^\prime}$ derived from 
three-body  rare $B$-decays  to dineutrinos (bottom row).
Numbers  correspond to a limit on the modulus except when a range is given.
   LFV-bounds are quoted as flavor-summed, $\sqrt{|\mathcal{K}^{\ell^+ \ell^{\prime-}}|^2 + |\mathcal{K}^{\ell^- \ell^{\prime +}}|^2}$. 
  }
  \label{tab:limits-13t}
\end{table}

The bounds \eqref{eq:tq}, \eqref{eq:tc} and  \eqref{eq:tu} imply upper limits on dineutrino branching ratios of FCNC top decays
\begin{align} \label{eq:top-br}
\mathcal{B}(t \to q \nu \bar \nu)=  \frac{G_F^2 m_t^5}{\Gamma_t 192 \pi^3} \left(\frac{\alpha_e}{4 \pi} \right)^2 x_{tq} \simeq  x_{tq} \, \cdot 10^{-9}  \, ,
\end{align}
which provide novel tests of LU and cLFC.\linebreak
$\Gamma_t=1.35\,\text{GeV}$ denotes the top width~\cite{Zyla:2020zbs}. 
LU \eqref{eq:tq} predicts hence 
 \begin{align} \label{eq:top-brLU}
 \mathcal{B}(t \to q \nu \bar \nu)_\text{LU} \lesssim 10^{-4}\, . 
 \end{align}
The bound is driven by the poorly known $\mathcal{K}_R^{{tq} \ell \ell} $ couplings, and much stronger ones are obtained 
assuming left-handed couplings only 
 \begin{align} 
 \mathcal{B}(t \to c \nu \bar \nu)_\text{LU}^\text{L} \lesssim 1\cdot 10^{-11}\, ,\label{eq:top-charm-brLU-left}\\ 
 \mathcal{B}(t \to u \nu \bar \nu)_\text{LU}^\text{L}  \lesssim 2\cdot 10^{-11}\, ,\label{eq:top-up-brLU-left}
 \end{align}
$SU(2)_L$ connects also $b\to s \nu \bar \nu$ and $b\to d \nu \bar \nu$ observables to
left-handed $tc$-- and $tu$--couplings with charged leptons, respectively. Corresponding limits are given in Tabs.~\ref{tab:limits-23t} and~\ref{tab:limits-13t}.

\subsection{Collider tests  with single tops \label{sec:1top}}

\begin{figure}[ht!]
    \centering
    \includegraphics[scale=0.8]{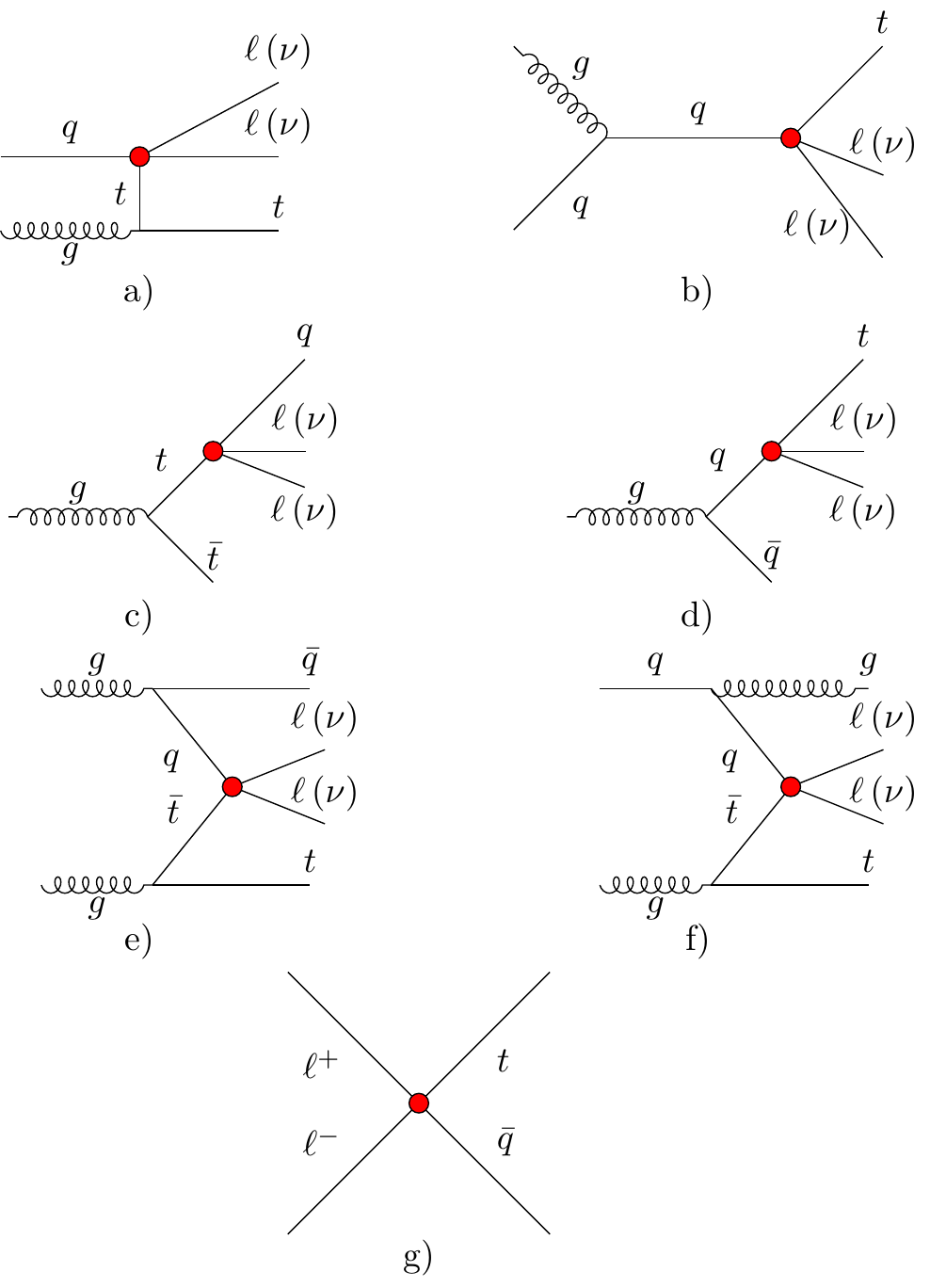}
   \caption{Diagrams with semileptonic four-fermion operators involving $qt\ell \ell^\prime$ or  $qt\nu  \bar \nu$, $q=u,c$ (red blobs)  that contribute to 
   a)+b): single top plus dileptons, or single top plus missing energy, wiggly line are gluons, c)-f):
   single top plus jet plus opposite sign dileptons, or  single top plus jet plus missing energy. Wiggly lines are gluons, suitable to hadron collider study, or electroweak gauge bosons
   $\gamma, Z$, which can also contribute at a lepton collider  (c)+d)). Contribution  g) matters only for lepton colliders, with $\ell^+ \ell^-$ annihilating into $t \bar q$.}
    \label{fig:collider}
\end{figure}

While the tests of LU and cLFC can be performed at the level of top branching ratios (\ref{eq:top-br}),  we propose tests with single top production
at the LHC, at least for the charged lepton couplings as in \cite{CMS:2021nlh,ATLAS:2018avw,CMS:2020lrr}, or a future  electron   \cite{Abada:2019lih}
or muon \cite{Delahaye:2019omf,Zimmermann:2018wfu} collider.
Semileptonic four-fermion operators involving $qt\ell \ell^\prime$ or  $qt\nu  \bar \nu$, $q=u,c$  contribute to
single top and single top plus jet signatures, shown in Fig.~\ref{fig:collider}. 

To begin let us recall that in the high energy limit all chiralities contribute without interference. In this limit,
constraints in top plus dilepton measurements appear in the following combination of Wilson coefficients
\begin{align}
|\mathcal{K}_L^{{tq} \ell \ell'}|^2+|\mathcal{K}_R^{{tq} \ell \ell'}|^2 +|\mathcal{G}_L^{{tq} \ell \ell'}|^2+|\mathcal{G}_R^{{tq} \ell \ell'}|^2  \, . 
\end{align}
Here the couplings 
$\mathcal{G}_{L,R}$ correspond to additional terms in the SMEFT Lagrangian~\cite{Grzadkowski:2010es}, with $SU(2)_L$-singlet leptons $E$
as
 $\frac{C_{qe}}{v^2} \bar Q \gamma_\mu Q \bar E \gamma^\mu E$  and  $\frac{C_{ue}}{v^2} \bar U \gamma_\mu U \bar E \gamma^\mu E$,
 where  $\mathcal{G}_L^{U,D}=(2 \pi/\alpha_e )C_{qe}$ and  $\mathcal{G}_R^{U}=(2 \pi/\alpha_e )C_{ue}$.
 The couplings to $SU(2)_L$-doublet quarks $\mathcal{G}_L^{{tq} \ell \ell'}$ are subjected to bounds on right-handed leptons from
$B$-decay via the weak effective theory coefficients $\sim C_9^{\ell \ell'}+C_{10}^{\ell \ell'}$, resulting in
$\mathcal{G}_L^{tc\mu \mu}=V_{tb}V_{ts}^*(-0.02\pm 0.49) \sim 0.02$, $\mathcal{G}_L^{tu\mu \mu}=V_{tb}V_{td}^*\,(-2\pm 6) \sim 0.1$~\cite{Bause:inprep}.

We illustrate the strategy for a test of LU in the top sector but note that the ingredients are analogous to the ones in charm, Sec.~\ref{sec:charm}.
Let us assume, for concreteness, that dimuons provide the best measurement of the cross section $\sigma_{\mu \mu}$ in single top plus 
opposite sign same flavor leptons.
It is induced by both left-handed and right-handed tops, $\sigma_{\mu \mu} =\kappa \sum_{q}\left( |\mathcal{K}_L^{{tq} \mu\mu}|^2+|\mathcal{K}_R^{{tq} \mu\mu}|^2 \right) +
\bar \sigma_{\mu \mu}$,
with proportonality constant $\kappa$.
We can use this expression to obtain from a measurement of $\sigma_{\mu \mu}$ an upper limit on  $|\mathcal{K}_R^{{tq} \mu\mu}|$; a stronger one would arise if 
the cross section induced by 
right-handed lepton contributions, $\bar \sigma_{\mu\mu} \propto |\mathcal{G}_L^{{tq} \ell \ell'}|^2+|\mathcal{G}_R^{{tq} \ell \ell'}|^2$, could be determined and subtracted, or $B$-physics constraints  on $|\mathcal{K}_L^{{tq} \mu\mu}|$ are imposed (see Tables~\ref{tab:limits-23t}-\ref{tab:limits-13t}). 
However,  the most conservative bound is obtained for $|\mathcal{K}_R^{{tq} \mu\mu}|$ saturating $\sigma_{\mu \mu}$.
Assuming LU, the single top plus dineutrinos cross section reads, using \eqref{eq:super},
\begin{align} \label{eq:full-LU}
\sigma_{\nu \nu}^{\text{LU}} =3 \kappa\left( \sum_{q=d,s} |\mathcal{K}_L^{{bq} \mu\mu}|^2+ \sum_{q=u,c} |\mathcal{K}_R^{{tq} \mu\mu}|^2 \right) \, , 
\end{align} hence
\begin{align} \label{eq:simple}
\sigma_{\nu \nu}^{\text{LU}}  \leq 3 \left( \sigma_{\mu \mu} + \kappa \sum_{q=d,s} |\mathcal{K}_L^{{bq} \mu\mu}|^2\right) \, , 
\end{align}
and $\mathcal{K}_L^{{bq} \mu\mu}$ is bounded by the global $b\to s$ and $b \to d$ fits, with results given in Tables~\ref{tab:limits-23b} and \ref{tab:limits-13b}.
To arrive at the simple expression (\ref{eq:simple}) we assumed kinematic cuts for dileptons and dineutrinos to be identical.
A violation of (\ref{eq:simple})  indicates breakdown of lepton universality.
Correspondingly, tests of cLFV are obtained as
\begin{align} \label{eq:full-cLFC}
\sigma_{\nu \nu}^{\text{cLFC}}  =\kappa \!\! \sum_{ \ell=e,\mu,\tau \!  \! }\left(  \sum_{q=d,s} |\mathcal{K}_L^{{bq} \ell\ell}|^2+ \sum_{q=u,c} |\mathcal{K}_R^{{tq} \ell \ell}|^2 \right) \, ,
\end{align}
hence 
\begin{align} \label{eq:simple2}
\sigma_{\nu \nu}^{\text{cLFC}}  \leq \sum_{ \ell=e,\mu,\tau}\left( \sigma_{\ell \ell} + \kappa \sum_{q=d,s} |\mathcal{K}_L^{{bq} \ell\ell}|^2\right) \, .
\end{align}
The expressions (\ref{eq:simple}), (\ref{eq:simple2})  are schematic only as SM contributions have not been taken into account. A full collider analysis is beyond the scope of this paper. Note also that SM contributions from $Z \to \ell^+ \ell^-$ could be controlled by cuts on the dilepon invariant mass, a feature that is not possible in
$Z \to \nu \bar \nu$, hampering dineutrino searches at the LHC.

A bound on lepton-flavor specific $qt\ell \ell$ operators from  $\ell^+ \ell^-$ annihilating into $t \bar q$ at a future lepton collider can be obtained.
The diagram for this process is shown  in Fig.~\ref{fig:collider}.

\begin{align}\sigma (\ell^+ \ell^- \to t \bar q)=\frac{G_F^2\,\alpha_e^2}{(4\pi)^3}\,s\, f(\xi)\,\left( |\mathcal{K}_L^{{tq} \ell \ell}|^2+|\mathcal{K}_R^{{tq} \ell \ell}|^2\right) + \bar \sigma
\end{align}
where $f(\xi)=\xi^2\,\,\big(1-\frac{1}{3}\xi\big)$ with $\xi=1-\frac{m_t^2}{s}$, and $\bar \sigma$ denotes contributions from  $SU(2)_L$-singlet leptons. In polarized $\ell^+ \ell^-$ collisions, the latter could be extracted.

We summarize strategies with single tops: Single top data in association with charged, opposite sign dileptons can be used to obtain quantitative predictions for
single tops with dineutrinos  assuming LU, or cLFC. Comparison
to requisite, actual measurements of single tops with dineutrinos, as in (\ref{eq:simple}), (\ref{eq:simple2}), allows to test the lepton flavor structure. The cLFC test requires
data on dielectron, dimuon and ditau spectra, whereas the LU test can be performed using dimuons, or more general, a single species' cross section, alone.
Irrespective of concrete studies with single tops associated with charged dileptons, the tests can be performed using  input on the SMEFT coefficients from elsewhere, as in 
(\ref{eq:full-LU}), (\ref{eq:full-cLFC}).  Presently, the right-hand sides are dominated by the poorly constrained couplings with right-handed tops, 
$\mathcal{K}_R^{{tq} \ell \ell}$. LHC sensitivities in single top production with dileptons \cite{Afik:2021jjh} are encouraging and suggest that the $3000 \, \text{fb}^{-1}$ HL-LHC can probe
couplings at the level of  $|\mathcal{K}_{R}^{tc\mu \mu}| \lesssim 22$
and $|\mathcal{K}_{R}^{tu\mu \mu}| \lesssim 2.2$. 
A dedicated analysis of missing energy collider distributions, {\it  i.e,} computing $\kappa$ as a function of kinematic cuts and efficiencies, would be desirable but is beyond the scope of this study.

\section{Summary \label{sec:con}}

Thanks to the flavor-inclusiveness
of missing energy measurements at particle physics experiments, $SU(2)_L$-links  of neutrinos with the charged leptons \eqref{eq:super} allow to probe lepton flavor structure  in dineutrino observables in three benchmarks: charged lepton flavor violation,
charged lepton flavor conservation and lepton universality.  
We put forward  concrete, novel tests  in charm, beauty and top, exploiting the connection between the up- and the down-sector.
Tests  invoke  experimental findings from the charged lepton sector and will evolve  with them in the future.

Key predictions for rare charm decays are  upper limits on the dineutrino branching ratios corresponding to the
benchmarks universality, lepton flavor conservation and including charged LFV, compiled in Table~\ref{tab:afactors}. 
The missing energy modes are well-suited for the experiments  Belle II \cite{Kou:2018nap}, BES III \cite{Ablikim:2019hff}, and future $e^+ e^-$-colliders, such as an FCC-ee running at the $Z$ \cite{Abada:2019lih} or the super-tau-charm factory (STCF)~\cite{Charm-TauFactory:2013cnj}, with sizable charm  rates. Since any observation of $c \to u \,\nu \bar \nu$ transitions heralds NP, experimental analysis is encouraged.

Interestingly, despite the lack of constraints  from top FCNCs, the beauty sector allows to test lepton universality  using dineutrino decays, thanks
 to correlations between $B$-decay modes involving different final state hadrons, notably $B \to K \nu \bar \nu$ versus
  $B \to K^* \nu \bar \nu$, and $B \to \pi  \nu \bar \nu$ versus
  $B \to \rho \nu \bar \nu$, see Sec.~\ref{sec:beauty}.  
   Corresponding tests are suitable  for Belle II and a $Z$-factory.
 
 Analyses  testing lepton flavor structure with rare top decays, discussed in Sec.~\ref{sec:top}, are similar to rare charm decays: a negligible SM background
  together with constraints on NP from the down-sector.
 In addition to FCNC top decays, rare top couplings from semileptonic four-fermion operators can be probed in single top plus dileptons   -- to be compared to 
single top plus missing energy, see Sec.~\ref{sec:1top}. While a dedicated sensitivity study is beyond the scope of this work we note that studies with ditops
and dileptons including $t q \ell \ell$ in the signal simulation are already available from the LHC \cite{CMS:2020lrr}.
Improving limits on such couplings is key to improve lepton flavor-symmetry predictions for $|\Delta t|= |\Delta (u,c)|=1$ dineutrino channels (\ref{eq:top-brLU}).

We conclude that processes with  dineutrinos  offer new and model-independent 
ways to test the SM and its approximate flavor symmetries, and  to shed light on the persistent hints for universality violation in $B$-decays, e.g., \cite{Bifani:2018zmi}.

 We close by commenting on the effects of light, right-handed neutrinos, not covered by the SMEFT framework, on the tests of lepton flavor structure
 \cite{Bause:2020xzj,Bause:2021ply}.
 The presence of both SM-like and right-handed neutrinos allows for  pseudo (-scalar)  four-fermion operators, that 
 induce decays of  mesons annihilating to dineutrinos.
 Quantitatively, improving the bound ${\mathcal{B}}(D^0 \to \nu \bar \nu) < 9.4 \cdot 10^{-5}$ at 90$\,\%$ CL~\cite{Lai:2016uvj} by about 2 orders of magnitude
 would suppress theses effects to be within the theoretical uncertainty in the $|\Delta c|= |\Delta u|=1$ studies reported here.
 Improving ${\mathcal{B}}(B_d^0 \to \nu \bar \nu) < 2.4  \cdot  10^{-5}$ at 90$\,\%$ CL~\cite{Zyla:2020zbs} down to $5 \cdot  10^{-7}$
 would suffice for the pseudo (-scalar) contribution to be at most percent level in semileptonic $|\Delta b|= |\Delta d|=1$ decays. Presently no bound on  $B^0_s \to \text{invisibles}$ exists but one at the level of Belle II projections,  ${\mathcal{B}}(B_s^0 \to \nu \bar \nu) < 1.1  \cdot  10^{-5}$ 
 with $0.5 \, \text{ab}^{-1}$~\cite{Kou:2018nap},would allow for similar control of the impact of right-handed neutrinos in $|\Delta b|= |\Delta s|=1$ decays. \\
 
 \textbf{Note added:} A very recent search by BES III reports ${\mathcal{B}}(D^0 \to \pi^0 \nu \bar \nu) < 2.1 \cdot 10^{-4}$ at 90$\,\%$ CL~\cite{BESIII:2021slf}, which is about  one  order of magnitude away from being constraining, see Table~\ref{tab:afactors}. 

\begin{acknowledgements}
This work is supported by the 
\textit{Studienstiftung des Deutschen Volkes} (MG) and the \textit{Bundesministerium f\"ur Bildung und Forschung} -- BMBF (HG).
\end{acknowledgements}


\begin{thebibliography}{10}

\bibitem{Grossman:1995gt}
Y.~Grossman, Z.~Ligeti and E.~Nardi,
Nucl. Phys. B \textbf{465}, 369-398 (1996)
[erratum: Nucl. Phys. B \textbf{480}, 753-754 (1996)]
doi:10.1016/0550-3213(96)00051-X
[arXiv:hep-ph/9510378 [hep-ph]].

\bibitem{Hiller:2014qzg}
G.~Hiller,
APS Physics \textbf{7}, 102 (2014).

\bibitem{Ciezarek:2017yzh}
G.~Ciezarek, M.~Franco Sevilla, B.~Hamilton, R.~Kowalewski, T.~Kuhr, V.~Lüth and Y.~Sato,
Nature \textbf{546} (2017), 227-233
[arXiv:1703.01766 [hep-ex]].

\bibitem{Eilam:1990zc}
G.~Eilam, J.~L.~Hewett and A.~Soni,
Phys. Rev. D \textbf{44}, 1473-1484 (1991)
[erratum: Phys. Rev. D \textbf{59}, 039901 (1999)]
doi:10.1103/PhysRevD.44.1473

\bibitem{Burdman:2001tf} 
  G.~Burdman, E.~Golowich, J.~L.~Hewett and S.~Pakvasa,
  Phys.\ Rev.\ D {\bf 66}, 014009 (2002)
  [hep-ph/0112235].

\bibitem{Grossman:2003rw}
Y.~Grossman, G.~Isidori and H.~Murayama,
Phys. Lett. B \textbf{588} (2004), 74-80
[arXiv:hep-ph/0311353 [hep-ph]].

\bibitem{Buras:2014fpa}
A.~J.~Buras, J.~Girrbach-Noe, C.~Niehoff and D.~M.~Straub,
JHEP \textbf{02} (2015), 184
[arXiv:1409.4557 [hep-ph]].

\bibitem{Grzadkowski:2010es}
B.~Grzadkowski, M.~Iskrzynski, M.~Misiak and J.~Rosiek,
JHEP \textbf{10} (2010), 085
[arXiv:1008.4884 [hep-ph]].

\bibitem{Efrati:2015eaa}
A.~Efrati, A.~Falkowski and Y.~Soreq,
JHEP \textbf{07} (2015), 018
[arXiv:1503.07872 [hep-ph]].

\bibitem{Brivio:2019ius}
I.~Brivio, S.~Bruggisser, F.~Maltoni, R.~Moutafis, T.~Plehn, E.~Vryonidou, S.~Westhoff and C.~Zhang,
JHEP \textbf{02} (2020), 131
[arXiv:1910.03606 [hep-ph]].

\bibitem{Alonso:2013hga}
R.~Alonso, E.~E.~Jenkins, A.~V.~Manohar and M.~Trott,
JHEP \textbf{04} (2014), 159
[arXiv:1312.2014 [hep-ph]].


\bibitem{Feruglio:2017rjo}
F.~Feruglio, P.~Paradisi and A.~Pattori,
JHEP \textbf{09}, 061 (2017)
doi:10.1007/JHEP09(2017)061
[arXiv:1705.00929 [hep-ph]].


\bibitem{Bause:2021ply}
R.~Bause, H.~Gisbert, M.~Golz and G.~Hiller,
JHEP \textbf{12}, 061 (2021)
doi:10.1007/JHEP12(2021)061
[arXiv:2109.01675 [hep-ph]].



\bibitem{Fuentes-Martin:2020lea}
J.~Fuentes-Martin, A.~Greljo, J.~Martin Camalich and J.~D.~Ruiz-Alvarez,
JHEP \textbf{11} (2020), 080.
[arXiv:2003.12421 [hep-ph]].

\bibitem{Angelescu:2020uug}
A.~Angelescu, D.~A.~Faroughy and O.~Sumensari,
Eur. Phys. J. C \textbf{80}, no.7, 641 (2020)
[arXiv:2002.05684 [hep-ph]].


\bibitem{Mandal:2019gff}
  R.~Mandal and A.~Pich,
  JHEP {\bf 1912} (2019) 089
  doi:10.1007/JHEP12(2019)089
  [arXiv:1908.11155 [hep-ph]].


\bibitem{Brod:2021hsj}
J.~Brod, M.~Gorbahn and E.~Stamou,
PoS \textbf{BEAUTY2020} (2021), 056
doi:10.22323/1.391.0056
[arXiv:2105.02868 [hep-ph]].

\bibitem{Zyla:2020zbs}
P.A.~Zyla \textit{et al.} [Particle Data Group],
PTEP \textbf{2020}, no.8, 083C01 (2020)
doi:10.1093/ptep/ptaa104
 
 
\bibitem{Bause:2019vpr}
R.~Bause, M.~Golz, G.~Hiller and A.~Tayduganov,
Eur. Phys. J. C \textbf{80} (2020) no.1, 65
[erratum: Eur. Phys. J. C \textbf{81} (2021) no.3, 219]
[arXiv:1909.11108 [hep-ph]].

\bibitem{Gisbert:2020vjx}
H.~Gisbert, M.~Golz and D.~S.~Mitzel,
Mod. Phys. Lett. A \textbf{36} (2021) no.04, 2130002
[arXiv:2011.09478 [hep-ph]].
 

\bibitem{Lai:2016uvj} 
  Y.-T.~Lai {\it et al.} [Belle Collaboration],
  Phys.\ Rev.\ D {\bf 95}, no. 1, 011102 (2017)
  [arXiv:1611.09455 [hep-ex]].

 
\bibitem{Bause:2020xzj}
R.~Bause, H.~Gisbert, M.~Golz and G.~Hiller,
Phys. Rev. D \textbf{103} (2021) no.1, 015033
[arXiv:2010.02225 [hep-ph]].







\bibitem{Bause:inprep}
R.~Bause, H.~Gisbert, M.~Golz and G.~Hiller, "Model-independent analysis of  $b  \to d$ processes",  DO-TH 21-30, 
\textit{in preparation}.




\bibitem{Kou:2018nap} 
  E.~Kou {\it et al.} [Belle-II Collaboration],
  PTEP {\bf 2019}, no. 12, 123C01 (2019)
  [arXiv:1808.10567 [hep-ex]].


\bibitem{CMS:2021nlh}
 [CMS],
``Search for charged lepton flavor violation in top quark production and decay in proton-proton collisions at $\sqrt{s}=13~\mathrm{TeV}$,''
CMS-PAS-TOP-19-006.

 
\bibitem{ATLAS:2018avw}
 [ATLAS],
``Search for charged lepton-flavour violation in top-quark decays at the LHC with the ATLAS detector,''
ATLAS-CONF-2018-044.
 


\bibitem{CMS:2020lrr}
A.~M.~Sirunyan \textit{et al.} [CMS],
JHEP \textbf{03}, 095 (2021)
doi:10.1007/JHEP03(2021)095
[arXiv:2012.04120 [hep-ex]].


\bibitem{Afik:2021jjh}
Y.~Afik, S.~Bar-Shalom, A.~Soni and J.~Wudka,
Phys. Rev. D \textbf{103}, no.7, 075031 (2021)
doi:10.1103/PhysRevD.103.075031
[arXiv:2101.05286 [hep-ph]].

\bibitem{Abada:2019lih} 
  A.~Abada {\it et al.} [FCC Collaboration],
  Eur.\ Phys.\ J.\ C {\bf 79}, no. 6, 474 (2019).




\bibitem{Delahaye:2019omf}
J.~P.~Delahaye, M.~Diemoz, K.~Long, B.~Mansouli\'e, N.~Pastrone, L.~Rivkin, D.~Schulte, A.~Skrinsky and A.~Wulzer,
[arXiv:1901.06150 [physics.acc-ph]].


\bibitem{Zimmermann:2018wfu}
F.~Zimmermann,
doi:10.18429/JACoW-IPAC2018-MOPMF065




  
\bibitem{Ablikim:2019hff}
M.~Ablikim {\it et al.},
Chin. Phys. C \textbf{44} (2020) no.4, 040001
[arXiv:1912.05983 [hep-ex]].
  
\bibitem{Charm-TauFactory:2013cnj}
A.~E.~Bondar \textit{et al.} [Charm-Tau Factory],
Phys. Atom. Nucl. \textbf{76} (2013), 1072-1085
doi:10.1134/S1063778813090032



\bibitem{Bifani:2018zmi}
S.~Bifani, S.~Descotes-Genon, A.~Romero Vidal and M.~H.~Schune,
J. Phys. G \textbf{46} (2019) no.2, 023001
[arXiv:1809.06229 [hep-ex]].


\bibitem{BESIII:2021slf}
M.~Ablikim \textit{et al.} [BESIII],
[arXiv:2112.14236 [hep-ex]].




\end{thebibliography}
\end{document}